\documentclass{article}
\usepackage{spconf,amsmath,epsfig}
\usepackage{url}
\usepackage{multirow}
\usepackage{threeparttable}
\usepackage{xcolor,colortbl}
\usepackage{subfigure}

\usepackage{amssymb}
\usepackage{amsmath}
\usepackage{subfiles}
\usepackage{makecell}
\usepackage{listings} 
\usepackage{adjustbox}
\usepackage{arydshln}
\usepackage{cancel}
\usepackage{multibib}
\usepackage{hyperref}
\newcites{SI}{Supplementary References}

\newcommand{\etal}{\textit{et al.}}
\newif\ifincludeSupplementaryOnly
\hypersetup{
    colorlinks=True,    
    pdfborder={0 0 0},   
}

\let\OLDthebibliography\thebibliography
\renewcommand\thebibliography[1]{
  \OLDthebibliography{#1}
  \setlength{\parskip}{0pt}
  \setlength{\itemsep}{0pt plus 0.3ex}
}

\begin{document}\sloppy

\def\x{{\mathbf x}}
\def\L{{\cal L}}
\ifincludeSupplementaryOnly
    \section*{\large Supplementary Information}

\setcounter{page}{1}
\setcounter{table}{0}
\setcounter{figure}{0}
\setcounter{footnote}{0}
\setcounter{section}{1}
This document provides more details about our proposed method and comparison.

\subsection{Detailed Architeture}

\subsubsection{Wavelet Up/Down Sampling}

For downsampling, features are first decomposed into four subbands, and their resolution is reduced by the DWT, followed by a 1$\times$1 convolution to reduce the channel dimension. For upsampling, on the other hand, they are first expanded back to their original channel dimensions and then we perform the IDWT, as shown in Fig.~\ref{fig:wave}.
\begin{figure}[h] 
\newcommand{\mywidth}{0.3}
\centering 
\subfigure[Wavelet Down]{\includegraphics[width=\mywidth\linewidth]{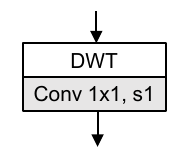}\label{subfig:down}} 
\hspace{1.5cm}
\subfigure[Wavelet Up]{\includegraphics[width=\mywidth\linewidth]
{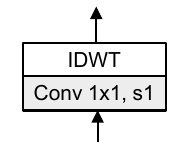}\label{subfig:up}}

\caption{\textbf{The structure of Wavelet Up/Down sampling.}
    }
\label{fig:wave} 
\end{figure}

\subsubsection{Cross-Attention}
The cross-attention module~\cite{chen2021crossvit} is illustrated in Fig.~\ref{subfig:attn_b}, and the attention mechanism is detailed in Fig.~\ref{subfig:cro}. This module~\cite{chen2021crossvit} is designed to harness low-resolution global information, thereby facilitating a more comprehensive global view. $F_H$ represents high-resolution features and $F_L$ denotes low-resolution features from the previous stage. CPE (conditional positional encoding) is utilized~\citeSI{chu2021conditional}.

\begin{figure}[h] 
\centering 
\subfigure[Cross Attention]{\includegraphics[width=0.27\linewidth]{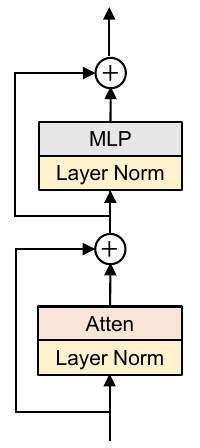}\label{subfig:attn_b}} 
\hspace{0.5cm}
\subfigure[Attention mechanism]{\includegraphics[width=0.45\linewidth]
{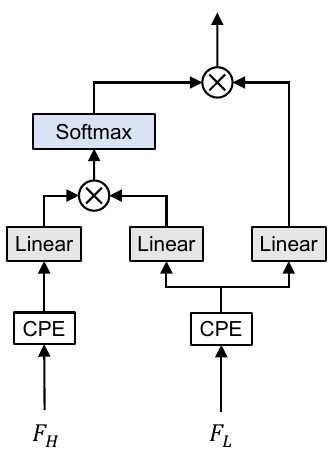}\label{subfig:cro}}

    \caption{\textbf{The structure of Cross Attention block.}} 
    \label{fig:cross} 
\end{figure}

\subsubsection{Latent Variable Block}
\label{sec:upper}
The primary distinction between the theoretical bound and the practical codec lies in their prior and posterior distributions. Specifically, in the practical codec's latent variable block, the posterior branch predicts only $\mu_i$. In contrast, the bound model's posterior branch predicts both $\mu_i$ and $\sigma_i$. Figure~\ref{fig:latents} provides a detailed view of the latent variable block. The state of the practical codec latent variable block during training is shown in Figure~\ref{subfig:train}. Figures~\ref{subfig:com} and \ref{subfig:decom} depict the encoding and decoding processes, respectively, while Figure~\ref{subfig:bound} displays the latent variable block for the bound model.

\textbf{Posteriors.} The posterior distribution of $Z_i$ given $x$ and $z<i$ in the practical model is:

\begin{equation}
\label{eq:post_s}
\begin{aligned}
q_i &\triangleq U(\mu_i - \frac{1}{2}, \mu_i + \frac{1}{2})
\\
\Leftrightarrow q_i(z_i \mid z_{<i},x) &= 
\left\{\begin{matrix}
1 & \text{for} \ \left | z_i - \mu_i \right | \leq \frac{1}{2} \\ 
0 & \text{otherwise}
\end{matrix}\right. ,
\end{aligned}
\end{equation}

where $\mu_i$ is the output of the posterior branch, depending on the image $x$ and preceding latent variables $ z_{<i}$. Once $q_i$ is obtained, $z_i$ is sampled as $z_i \leftarrow \mu_i + u$, where $u$ is a random sample from $U(-\frac{1}{2}, \frac{1}{2})$ during training, and during testing, it is replaced with scalar quantization.

In the theoretical bound model, the posterior distribution is a conditional Gaussian:
\begin{equation}
\label{eq:post_t}
\begin{aligned}
q_i &\triangleq \mathcal{N}({\mu}_i, {\sigma}_i^2) 
\end{aligned}
\end{equation}

\textbf{Priors.} For the practical model, the prior distribution $p_i$ is defined as a conditional Gaussian convolved with a uniform distribution:

\begin{equation}
\label{eq:prior_s}
\begin{aligned}
p_i &\triangleq \mathcal{N}(\hat{\mu}_i, \hat{\sigma}_i^2) * U(- \frac{1}{2}, \frac{1}{2})
\\
\Leftrightarrow p_i(z_i \mid z_{<i}) &= \int_{z_i-\frac{1}{2}}^{z_i+\frac{1}{2}} \mathcal{N}(t; \hat{\mu}_i, \hat{\sigma}_i^2) \ dt ,
\end{aligned}
\end{equation}
where $\mathcal{N}(t;\hat{\mu}_i, \hat{\sigma}_i^2)$ represents the Gaussian probability density function evaluated at $t$, and $t$ is an integration dummy variable. The mean $\hat{\mu}_i$ and standard deviation $\hat{\sigma}_i$ are predicted by the prior branch. The probability mass function (PMF) $P_i$ is then defined as:
\begin{equation}
P_i(n) \triangleq p_{i}(\hat{\mu}_i + n| z_{<i}), n \in \mathbb{Z}.
\end{equation}
which is used for the entropy coding/decoding of $z_i$.

The theoretical bound model also employs a conditional Gaussian as the prior distribution:
\begin{equation}
\label{eq:prior_t}
\begin{aligned}
p_i &\triangleq \mathcal{N}(\hat{\mu}_i, \hat{\sigma}_i^2) 
\end{aligned}
\end{equation}

\begin{figure*}[hthp] 
\newcommand{\mywidth}{0.45}
\centering 
\subfigure[Training]{\includegraphics[width=\mywidth\linewidth]{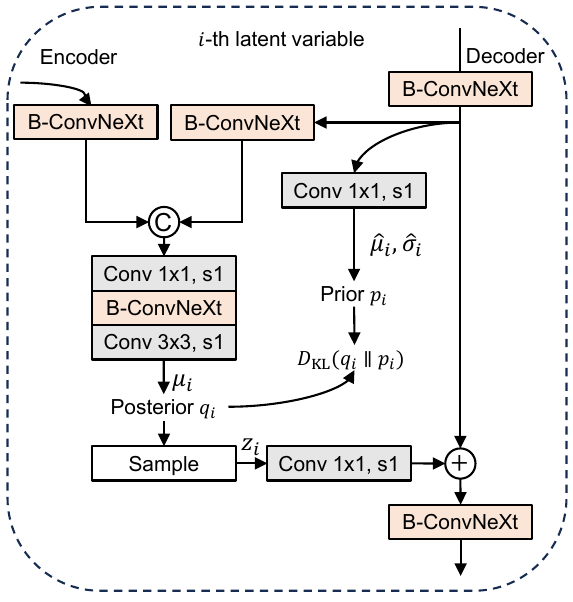}\label{subfig:train}} 
\subfigure[Bound]{\includegraphics[width=\mywidth\linewidth]{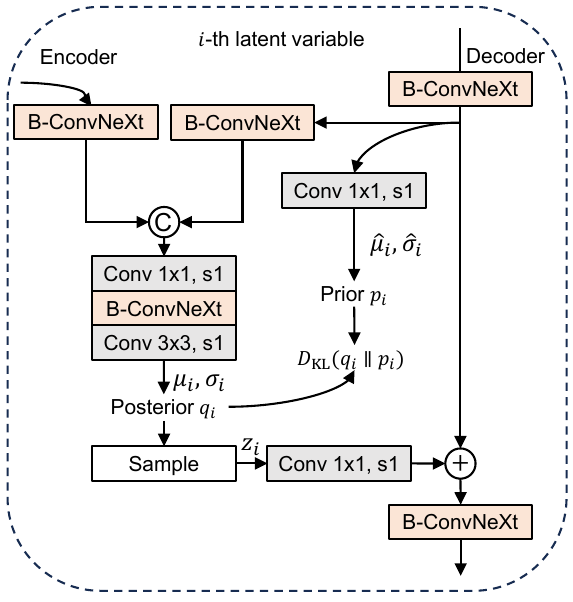}\label{subfig:bound}} 
\subfigure[Testing: encoding]{\includegraphics[width=\mywidth\linewidth]{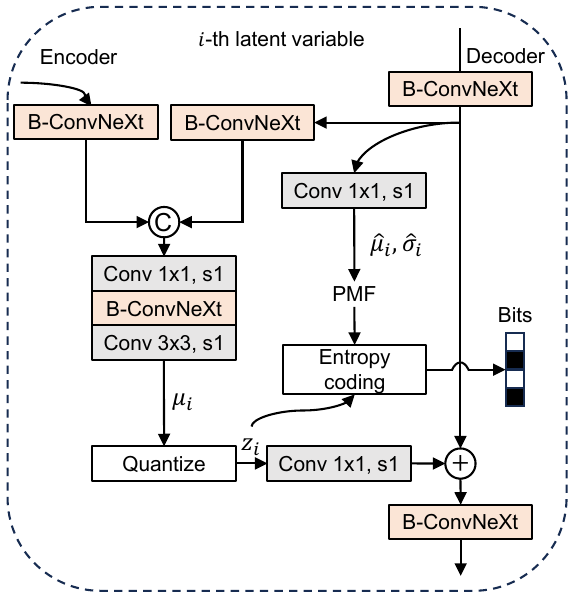}\label{subfig:com}}
\subfigure[Testing: decoding]{\includegraphics[width=\mywidth\linewidth]{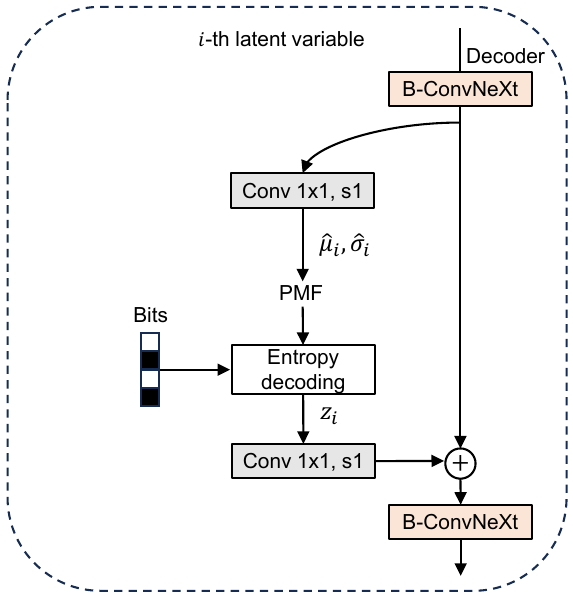}\label{subfig:decom}}
\caption{\textbf{Detailed Illustration of the Latent Variable Block.}
    }
\label{fig:latents} 
\end{figure*}

\subsection{Codecs Implementations}

\lstset{
  frame=shadowbox, 
  rulesepcolor=\color{red!20!green!20!blue!20},
  commentstyle=\color{red!10!green!70}\textit,    
  showstringspaces=false,
  numbers=left, 
  numberstyle=\tiny,    
  stringstyle=\ttfamily, 
  breaklines=true, 
  extendedchars=false,  
  texcl=true}

\subsubsection{Learning-based Codecs}

We list the implementations of the learning-based image codecs that we used for comparison in Table~\ref{tab:com_imp}.

\begin{table}[ht]
\centering
\footnotesize
\caption{Learning-based Codecs Implementations.}
\begin{adjustbox}{width=\linewidth}
\begin{tabular}{c|l}
\hline
\textbf{Method}                             & \multicolumn{1}{c}{\textbf{Implementation}} \\ \hline
M\&S Hyperprior~\cite{minnen2018joint}    & \url{github.com/InterDigitalInc/CompressAI}       \\ \hline
Cheng2020~\cite{cheng2020learned}       &\url{github.com/InterDigitalInc/CompressAI}       \\ \hline
STF~\cite{zou2022devil}&\url{github.com/Googolxx/STF}       \\ \hline
TCM-S~\cite{liu2023learned}&\url{github.com/jmliu206/LIC_TCM}       \\ \hline
QARV~\cite{duan2023qarv}&\url{github.com/duanzhiihao/lossy-vae}       \\ \hline

\end{tabular}
\end{adjustbox}
\label{tab:com_imp}
\end{table}

\subsubsection{VVC Codec}
We use VTM-18.0\footnote{\url{https://vcgit.hhi.fraunhofer.de/jvet/VVCSoftware_VTM/-/tree/VTM-18.0?ref_type=tags}}, the reference software for VVC, as the anchor. When testing VTM-18.0, we use OpenCV\footnote{\url{https://github.com/opencv/opencv/tree/4.8.0}} to convert the image to YUV444 format and then use the following command line to test it.
The output image is in YUV 4:4:4 format and converted back to RGB space. The final PSNR is computed between the final RGB image and the original RGB image.

\begin{lstlisting}
EncoderApp.exe 
        -c encoder_intra_vtm.cfg       
        -o output.yuv
        -q qp 
        -wdt image width 
        -hgt image height 
        -i input.yuv   
        -f 1 
        -fr 1 
        -fs 0 
        -b output.bin
        --InputChromaFormat=444
\end{lstlisting}

\subsection{Detailed Training Settings}

We list detailed training information in Table~\ref{table:hyp_param}, including data augmentation, hyperparameters, and training devices.
We use different settings for the main experiment and the ablation experiments.
In the main experiment, we train our model until convergence, which requires around 10 days of training on a dual-GPU machine.
For ablation study experiments, we train our model with a shorter training period (500k iterations instead of 2M iterations) to reduce training costs.

\begin{table}[htbp]
\small
\centering
\caption{Training Hyperparameters.}
\begin{tabular}{l|c|c}
               & Main model         & Ablation study      \\ \hline
Training set   & COCO 2017 train    & COCO 2017 train     \\
\# images      & 118,287            & 118,287             \\
Image size     & Around 640x420     & Around 640x420      \\ \hdashline
Data augment.  & Crop, h-flip       & Crop, h-flip        \\
Train input size & 256x256          & 256x256             \\ \hdashline
Optimizer      & Adam               & Adam                \\
Learning rate  & $2 \times 10^{-4}$ & $2 \times 10^{-4}$  \\
LR schedule    & Constant + cosine  & Constant            \\ \hdashline
Batch size     & 32                 & 32                  \\
\# iterations  & 2M                 & 500K                \\
\# images seen & 64M                & 16M                  \\ \hdashline
Gradient clip  & 2.0                & 2.0                 \\
EMA            & 0.9999             & 0.9999              \\ \hdashline
GPUs           & 2 $\times$ RTX 3090 & 1 $\times$ Quadro 6000 \\
Time           & 260h               & 87h                 \\
\end{tabular}
\label{table:hyp_param}
\end{table}

\subsection{RD curves Magnified}
RD curves are magnified in Figures~\ref{fig:sup_kodak}, \ref{fig:sup_tecnick}, and \ref{fig:sup_CLIC} for the convenience of observation.

\begin{figure*}[htbp] 
    \centering 
    \includegraphics[width=\linewidth]{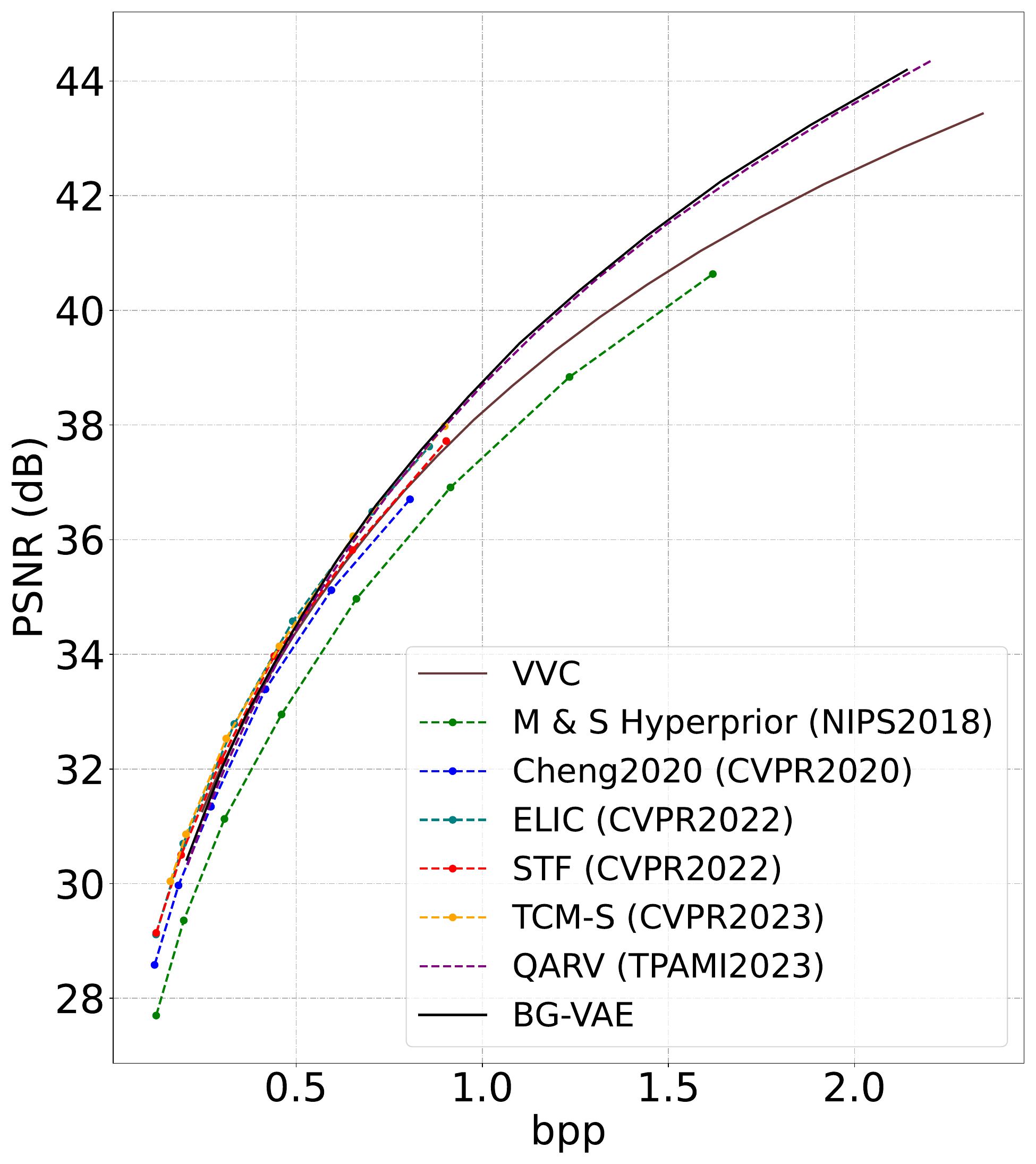}
    \caption{{RD Curves on Kodak Dataset}.} 
    \label{fig:sup_kodak} 
\end{figure*}

\begin{figure*}[htbp] 
    \centering 
    \includegraphics[width=\linewidth]{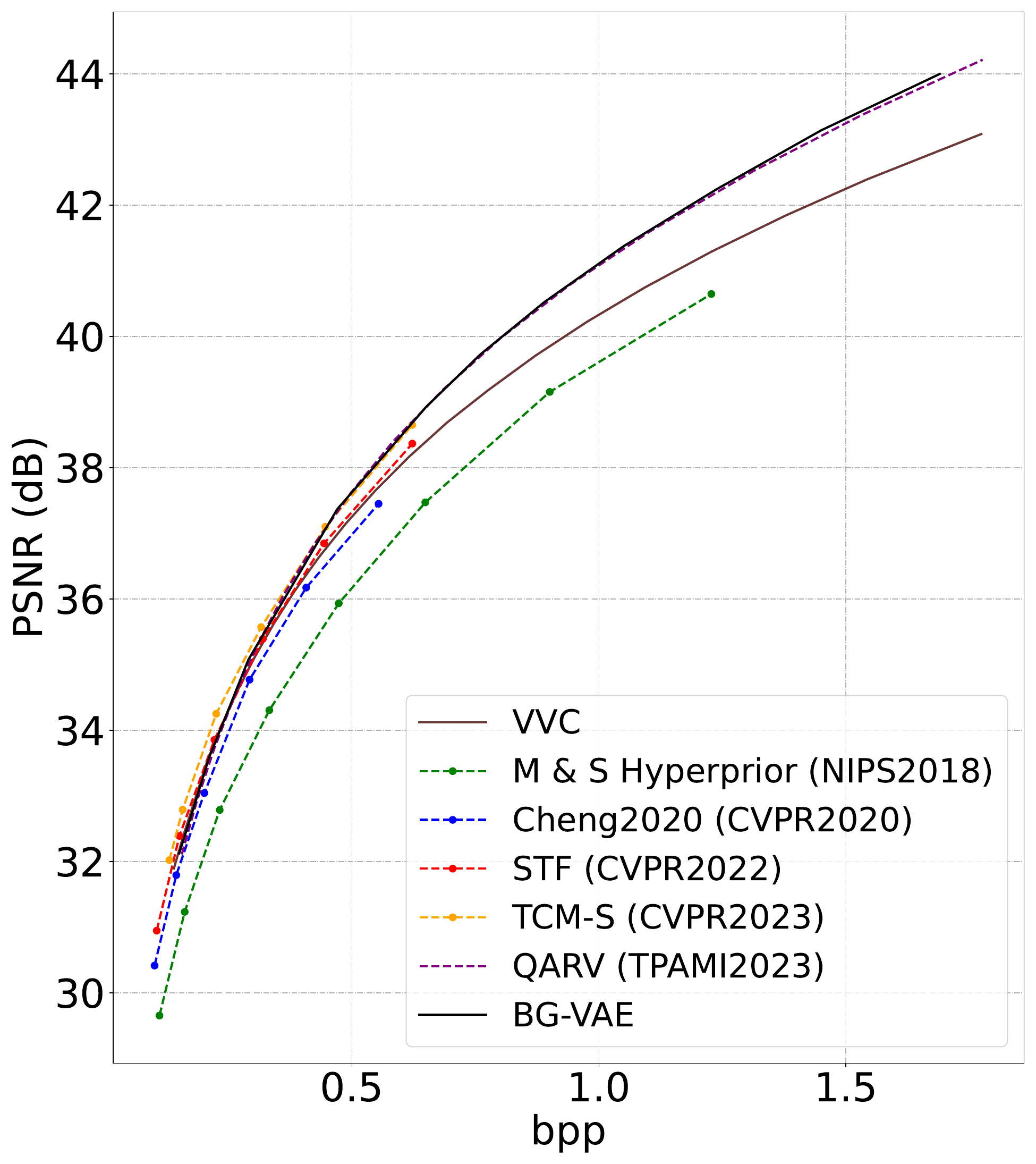}
    \caption{{RD Curves on Tecnick Dataset}.} 
    \label{fig:sup_tecnick} 
\end{figure*}

\begin{figure*}[htbp] 
    \centering 
    \includegraphics[width=\linewidth]{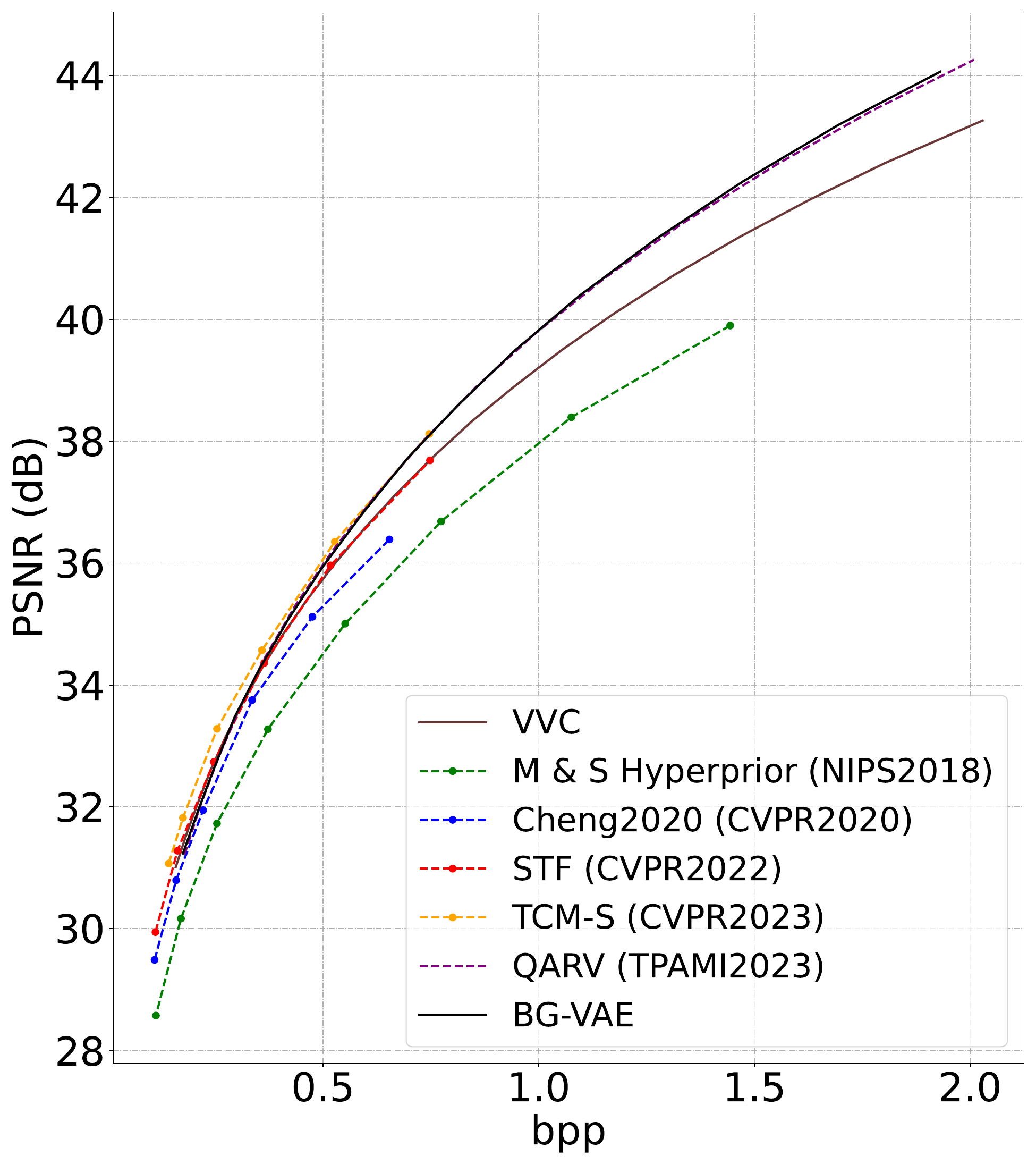}
    \caption{{RD Curves on CLIC2022 Dataset}.} 
    \label{fig:sup_CLIC} 
\end{figure*}

\bibliographystyleSI{IEEEbib}
\bibliographySI{QARV2}
\else


\title{Theoretical Bound-Guided Hierarchical VAE for Neural Image Codecs}

%
\name{Yichi Zhang, Zhihao Duan, Yuning Huang, Fengqing Zhu}
\address{Elmore Family School of Electrical and Computer Engineering,\\Purdue University, West Lafayette, Indiana, U.S.A.\\\{zhan5096, duan90, huan1781, zhu0\}@purdue.edu}

\maketitle

\begin{abstract}
\vspace{-0.2cm}

Recent studies reveal a significant theoretical link between variational autoencoders (VAEs) and rate-distortion theory, notably in utilizing VAEs to estimate the theoretical upper bound of the information rate-distortion function of images.
Such estimated theoretical bounds substantially exceed the performance of existing neural image codecs (NICs).
To narrow this gap, we propose a theoretical bound-guided hierarchical VAE (BG-VAE) for NIC.
The proposed BG-VAE leverages the theoretical bound to guide the NIC model towards enhanced performance.
We implement the BG-VAE using Hierarchical VAEs and demonstrate its effectiveness through extensive experiments.
Along with advanced neural network blocks, we provide a versatile, variable-rate NIC that outperforms existing methods when considering both rate-distortion performance and computational complexity.
The code is available at \href{https://gitlab.com/viper-purdue/BG-VAE}{BG-VAE}.

\end{abstract}

\begin{keywords}
Lossy Image Compression, Knowledge Distillation, Hierarchical VAE
\end{keywords}

\vspace{-0.3cm}
\section{Introduction}
\label{sec:intro}
\vspace{-0.3cm}

Lossy image compression, an important problem in image processing, aims to compress images to a low-rate representation while retaining high reconstruction quality.
It is essential for efficient storage and transmission in numerous applications, from web media to satellite imagery.
With the rapid advances in deep learning, the landscape of lossy image compression has undergone a significant transformation. Recent works not only achieved substantial progress in practical compression methods but also deepened the theoretical analysis.
The majority of research in this field focuses on the practical aspect, where various neural network architectures~\cite{zou2022devil,he2022elic} and probability models~\cite{cheng2020learned,duan2023qarv} have been proposed to improve the rate-distortion (R-D) performance.
This leads to a series of strong neural image codecs (NICs)~\cite{he2022elic,liu2023learned,zhang2024Another} that outperform traditional codecs such as H.266/VVC~\cite{bross2021overview}.

Despite the impressive R-D performance of recent NICs, it has been shown that there is still a considerable gap between their performance and theoretically achievable limits~\cite{yang2022sandwich, duan2023improved}.
Recall that the best achievable R-D performance for a data source is described by its \textit{information rate-distortion (R-D) function}, a fundamental quantity in lossy compression.
By bounding the information R-D function using deep variational autoencoders~\cite{kingma2013auto} (VAEs), Yang and Mandt~\cite{yang2022sandwich} show that NICs could be improved by at least +1 dB in PSNR at various rates.
Interestingly, the neural network models used for bounding the information R-D function closely resemble recent NICs in the sense that both are hierarchical VAEs.

Inspired by this resemblance, this paper explores the possibility of leveraging the estimated bound information R-D functions to improve the NICs.
Specifically, we leverage the teacher-student framework~\cite{gou2021knowledge} that is prevalent in knowledge distillation methods.
The NIC's theoretical bound naturally serves as a \textit{teacher}, while the NIC itself takes the role of the \textit{student}.
During training, the NIC is guided by a theoretical bound model, thereby enhancing its efficacy.
The similarity in model structure and model size between the bound and the NIC obviates the need for a typically larger teacher model in the teacher-student framework.
The availability of theoretical bound simplifies the teacher selection process and reduces training resource consumption.
Furthermore, we exemplify a hierarchical NIC named BG-VAE (theoretical Bound-Guided VAE) with well-designed modules. Extensive experiments demonstrate the superior performance of our method compared to existing approaches.


Our contributions are summarized as follows:
\begin{itemize}
\vspace{-0.2cm}
\item {We propose a teacher-student framework that uses theoretical bounds to guide the training of NICs.}
\vspace{-0.2cm}
\item {We present new network modules and construct an efficient hierarchical model that leverages spatial and spectral information.}
\vspace{-0.2cm}
\item {Overall, we present BG-VAE, a NIC framework including the teacher-student training strategy and model architectures. Extensive experiments are conducted to demonstrate the effectiveness of the proposed method.}
\end{itemize}

\vspace{-0.55cm}

\section{Related work}
\vspace{-0.3cm}

In this section, we briefly review previous works and summarize the preliminaries.
\vspace{-0.3cm}

\subsection{Neural Image Codecs and VAEs}
\vspace{-0.2cm}

Most existing NICs follow the scheme of \textit{transform coding}, where images are transformed to a latent space for de-correlation and energy compression, followed by quantization and entropy coding. Early works stack convolutions to parameterize the codec~\cite{balle2016end, balle2018variational}. Later, attention mechanisms~\cite{cheng2020learned} and transformers~\cite{lu2022high,zhang2024reconfigurable}, are developed to improve the performance. Another line of research focuses on developing context models to aid entropy coding involved with leveraging hierarchical latent variables~\cite{balle2018variational,hu2020coarse,duan2023qarv} and exploring correlations between pixels~\cite{minnen2018joint} as well as between channels of latent variables~\cite{he2022elic}. Essentially, the framework of these methods is similar to a one-layer~\cite{balle2016end} or two-layer~\cite{balle2018variational} VAEs.

VAEs~\cite{kingma2013auto} is a critical class of latent variable models, especially effective for complex data like images. In VAEs, data $X$ and latent variable $Z$ are linked through the joint distribution $p_{X, Z}(x, z) = p_{X|Z}(x|z) \cdot p_Z(z)$. VAEs feature an encoder-like approximate posterior $q_{Z|X}$ and a decoder-like $p_{X|Z}$, framing them as stochastic autoencoders.

In the NIC scenarios, VAEs employ a deterministic decoder $f_\text{dec}$, leading to a lossy reconstruction $\hat{X} = f_\text{dec}(Z)$. For high-dimensional data, hierarchical VAEs~\cite{kingma2016iafvae} enhance flexibility and expressiveness. They employ a series of latent variables $Z_{1:N} \triangleq \{Z_1, ..., Z_N\}$ in an autoregressive manner: $p_{Z_{1:N}} = \prod_{i=1}^{N} p_{Z_i|Z_{<i}}$. The architecture progresses from low to high dimensions, capturing the image's granular details.
The hierarchical VAE-based NIC training loss extends the standard single-layer loss across multiple latent variables:
\begin{equation}
\label{eq:lossrd}
    \mathcal{L}_\lambda = \mathbb{E}_{X \sim p_\text{data}, Z_{1:N} \sim q_{1:N}} \left[ \sum_{i=1}^N D_\text{KL}(q_i || \, p_i) + \lambda \cdot d(X,\hat{X}) \right],
\end{equation}
$q_i$ and $p_i$ denote the posterior and prior of each latent variable, approximated via ancestral sampling.

As for its theoretical bounds, Yang \etal~\cite{yang2022sandwich} established an upper bound on the information R-D function for image sources by training VAEs. In doing so, they show that there still exists a sizeable room for improvement over current NICs (+1 dB in PSNR at various rates). Duan \etal~\cite{duan2023improved} reported an improved upper bound on the information R-D function of images by extending it to hierarchical VAE and variable rate image compression, which showed that at least 30\% BD-rate reduction w.r.t. the VVC codec is achievable.

\vspace{-0.3cm}
\subsection{Knowledge Distillation}
\vspace{-0.2cm}

Knowledge Distillation (KD), initially introduced by Hinton \etal~\cite{hinton2015distilling}, balances model performance and efficiency. It serves as a training strategy, enhancing the performance of lightweight networks through knowledge transfer from a high-capacity teacher model, predominantly used in image classification~\cite{hinton2015distilling} and object detection~\cite{gou2021knowledge}. 
In contrast, KD in NIC remains relatively unexplored. Fu \etal~\cite{fu2023fast} developed an improved three-step KD training scheme for balancing decoder network complexity and performance, transferring both final and intermediate outputs from the teacher to the student network. However, this approach still relies on a large teacher model and a less effective MSE loss function.

{\bf Remarks.} Prior efforts in NIC predominantly emphasize model designs to achieve commendable performance. However, another effective way may involve identifying the theoretical bound of NIC and utilizing the bound to improve practical performance. Our BG-VAE begins with bound-guided training, forcing the practical NIC (\textit{student}) to mimic the bound (\textit{teacher}) behavior. Due to the similarity in model structure and model size of the bound and NIC, this method bypasses the need for a larger teacher model, as is customary in the KD method, thereby reducing training complexity.

\vspace{-0.3cm}
\section{Proposed Method}
\vspace{-0.3cm}

An overview of the proposed BG-VAE is presented in Fig.~\ref{fig:Overview}.  We begin by describing the bound-guided training and the loss function, as elaborated in Section~\ref{sec:distill}. Subsequently, the neural network architecture employed for implementing our BG-VAE is described in Section~\ref{sec:archi}.


\begin{figure*}[t] 
\centering 
\subfigure[Teacher-student training]{\raisebox{0.35cm}{\includegraphics[width=0.32\linewidth]{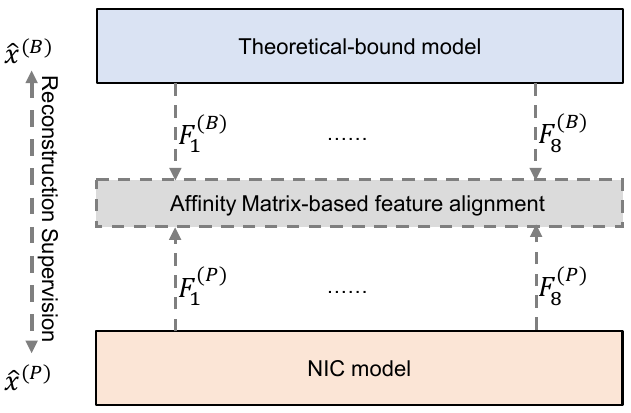}}\label{subfig:BG}}
\hspace{0.1cm}
\subfigure[Model architecture]{\includegraphics[width=0.65\linewidth]{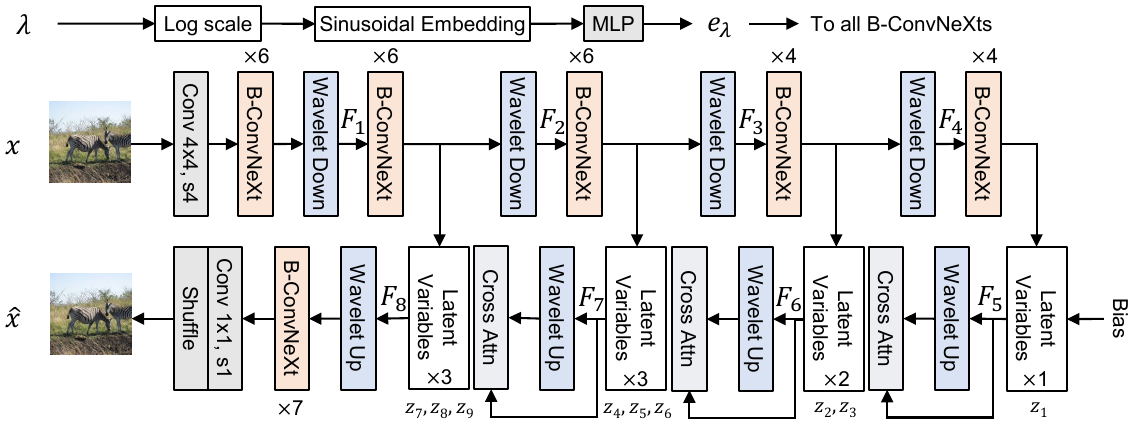}\label{subfig:model}}
\vspace{-0.3cm}
\caption{\textbf{An overview of the proposed BG-VAE.} (a) the bound-guided framework, (b) the model used to implement BG-VAE.}
\vspace{-0.5cm}
\label{fig:Overview} 
\end{figure*}

\vspace{-0.3cm}
\subsection{Theoretical Bound-Guided Training}
\vspace{-0.2cm}

\label{sec:distill}

{
As depicted in Fig.~\ref{subfig:BG}, our training recipe involves two models: a model that is used for estimating theoretical bounds, denoted by $B$, and the NIC model for practical deployment, denoted by $P$.
We refer to the former model as the \textit{teacher} and the latter as \textit{student}.
}
The teacher model's implementation adheres to the approaches from~\cite{duan2023improved}.
Both models share a similar network structure as shown in Fig~\ref{subfig:model}, and \( P \) is trained to mimic \( B \)'s behavior including the intermediate features and the reconstructed image.

\vspace{-0.3cm}
\subsubsection{Affinity Matrix-based Feature Alignment}
\vspace{-0.2cm}

Precise similarity measurement is crucial in feature alignment, especially given the unbounded feature representation space in regression problems~\cite{he2020fakd}, a similar case for NICs. This contrasts with the more constrained feature spaces in classification problems. As a result, distillation methods effective in classification~\cite{hinton2015distilling} are not suitable for compression. Inspired by FAKD~\cite{he2020fakd}, we develop an affinity matrix-based feature alignment process.

We extract feature maps at two points: after downsampling ({\textit{Wavelet Down} in Fig.~\ref{subfig:model}}) and before the upsampling ({\textit{Wavelet Up} in Fig.~\ref{subfig:model}}). For the teacher model, the feature map \( F^{(B)} \in \mathbb{R}^{b \times c \times \frac{w}{n} \times \frac{h}{n}} \) is extracted directly, with \( b \), \( c \), \( w \), \( h \), \( n \), represent the batch size, channels, width, height, the downsampled ratio at that stage, respectively. For the student model, we first use two B-ConvNeXt blocks (refer to Sections~\ref{sec:archi}) to transform the feature map, and subsequently, we get $F^{(P)}\in \mathbb{R}^{b\times c\times \frac{w}{n}\times \frac{h}{n}}$. Notably, we employ an ensemble strategy~\cite{chen2022improved} for the two B-ConvNeXt blocks.

Next, we compute the affinity matrix ${A}^{(B)}, {A}^{(P)}$ using $F^{(B)}$ and $F^{(P)}$. Our method emphasizes the spatial relationships within ${A}^{\{(B), (P)\}} \triangleq A$.  This is achieved by normalizing the matrix along the channel dimension and excluding this dimension through matrix multiplication during the computation. The affinity matrix $A$ is thus formulated by considering the spatial interrelations between pixels, highlighting their mutual dependencies and interactions. The affinity matrix $ A\in \mathbb{R}^{b\times \frac{hw}{n^2}\times \frac{hw}{n^2}}$ is determined by:
\begin{equation}
{A} = \left( \frac{F}{\sqrt{\sum_{k=1}^{c} F_k^2} + \varepsilon} \right)^T \times \frac{F}{\sqrt{\sum_{k=1}^{c} F_k^2} + \varepsilon}
\end{equation}
where $F$ is reshaped to $b\times c\times \frac{hw}{n^2}$ first. $F_k$ is the $k$-th channel of $F$. $\varepsilon$ is 1e-8 used to improve numerical stability.

To assess the similarity between $A^{(B)}$ and $A^{(P)}$, we utilize two indicators, the L1 loss for element-by-element difference assessment, and the cosine similarity loss, which evaluates the angular relationship between matrices, offering insights beyond mere element-wise comparison. Consequently, the feature affinity-based guidance loss \( \mathcal{L}_{\text{feature}} \) is expressed as:

\begin{equation}
\mathcal{L}_\text{feature} = \underbrace{\frac{1}{n} \sum_{i=1}^{n} |A^{(B)}_i - A^{(P)}_i|}_{\text{L1 Loss}} + \underbrace{1 - \frac{A^{(B)} \cdot A^{(P)}}{\|A^{(B)}\| \|A^{(P)}\|}}_{\text{Cosine Similarity Loss}}
\end{equation}

where $A^{(B)}_i$ and  $A^{(P)}_i$ are the $i$-th element of the teacher and student model's affinity matrix respectively.
\vspace{-0.3cm}
\subsubsection{Overall Loss}
\label{sec:over}
\vspace{-0.2cm}

Our approach includes the teacher model's supervisory role over the output of the practical codec. This is achieved by calculating the reconstruction supervision loss, denoted as \( \mathcal{L}_{\text{rs}} \):

\begin{equation}
\mathcal{L}_\text{rs} ={\frac{1}{n} \sum_{i=1}^{n} \left| \hat{x}_{i}^{(B)} - \hat{x}_{i}^{(P)}) \right|
}
\end{equation}

$\hat{x}_{i}^{(B)}$ and $\hat{x}_{i}^{(P)}$ are the $i$-th elements of the reconstructed image from the teacher model and the student model, respectively. Consequently, the final loss function is as follows:
\begin{equation}
\label{eq:loss_final}
    \mathcal{L} = \mathcal{L}_\lambda+w_1\cdot \sum_{j=1}^{8}{\mathcal{L}_{\text{feature}_j}}+ w_2\cdot\mathcal{L}_\text{rs} 
\end{equation}
where $w_1$ and $w_2$ are weighting factors adjusting magnitude orders which are set to 1 in our method. $\mathcal{L}_{\text{feature}_j}$ represents the loss calculated by using $F_{j}$. Variable-rate training is achieved by $\mathcal{L}_\lambda$ (Eq.~\eqref{eq:lossrd}), where $\lambda$ is randomly sampled from the range $[64, 8192]$ during training (this range approximately corresponds to $[0.2, 2.2]$ bits per pixel (bpp)).
{The entire model is conditioned on $\lambda$}
using the $\lambda$-embedding network (shown at the top of Fig.~\ref{subfig:model}) and through B-ConvNeXt.

{
At test time, the coding rate can be adjusted by tuning the input $\lambda$, effectively achieving variable-rate compression.
}

\vspace{-0.3cm}
\subsection{Model Architectures}
\vspace{-0.2cm}

\label{sec:archi}
This section presents a neural network implementation of the BG-VAE framework and its components in detail. 

{As shown in Fig~\ref{subfig:model}, to build the BG-VAE, we utilized a hierarchical VAE-based structure transmitting 9 latent variables ($z_1$ to $z_9$) in four stages. In the top-to-bottom path, ratio 4 downsampling is first achieved using a 4$\times$4, stride 4 convolution, followed by feature extraction using B-ConvNeXt and Wavelet Downsampling for ratio 2 downsampling. Latent Variable Blocks transmit latent variables $z_i$. In the bottom-to-top path, Wavelet Upsampling is employed for ratio 2 upsampling. To improve global information integration, we integrate a cross-attention module~\cite{chen2021crossvit}, which harnesses low-resolution global information, thereby facilitating a more comprehensive global view. Finally, 1$\times$1, stride 1 convolution, and shuffle enable ratio 4 upsampling. Each module will be elaborated in the following.}

\textbf{Balanced ConvNeXt block.} Based on ConvNeXt block~\cite{wang2022anti,liu2022convnet}, we introduce the Balanced ConvNeXt block (B-ConvNeXt), depicted in Fig.~\ref{fig:resb}. While retaining the core architecture of ConvNeXt, we implement key modifications: (1): After depth-wise convolution, the DC (direct current component) is extracted by averaging each channel and subtracted from the original features to isolate the HC (high-frequency component). Two learnable parameters $\alpha, \beta$ are then applied to modulate the balance between the DC and HC before re-integrating them with the original features; (2): We enable variable rates by conditioning on \(e_{\lambda}\), an embedded output from the \(\lambda\) embedding network (illustrated at the top of Fig.~\ref{subfig:model}). $e_\lambda$ passes through the GELU function and a Linear layer to scale features after LayerNorm in the block, making them conditional on $\lambda$.

\begin{figure}[htbp] 
    \centering 
    \includegraphics[width=0.7\linewidth]{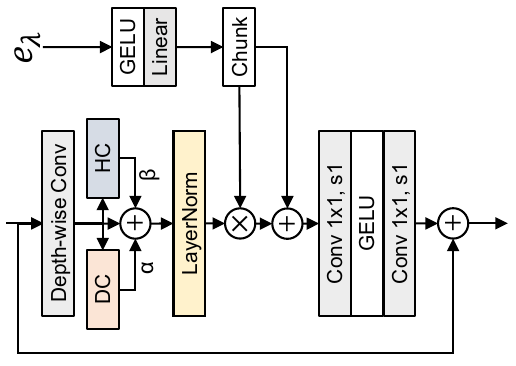}
    \vspace{-0.5cm}
    \caption{\textbf{The structure of Balanced ConvNeXt block.}} 
    \label{fig:resb} 
    \vspace{-0.5cm}
\end{figure}

\textbf{Wavelet Up/Down Sampling.} We utilize the Discrete Wavelet Transform (DWT)  for up/down sampling. DWT divides data into four subbands: the ILL, capturing coarse object structures, and the ILH, IHL, and IHH, detailing fine textures. These subbands are merged using a single convolution, reducing dimensions. Before applying the inverse DWT (IDWT), another single convolution is used to re-establish the original dimensions. DWT and IDWT are {invertible operations}, which preserves feature quality and fidelity.



\textbf{Latent Variable Blocks.} The Latent Variable Block is shown in Fig.~\ref{fig:latent}. The left-hand side of Fig.~\ref{fig:latent} shows the posterior branch of the latent variable, and the middle part indicates the prior branch. The posterior branch consists of three B-ConvNeXt blocks, a concatenation operation, and two convolutions. The prior branch, in contrast, consists of a single convolution that facilitates decoding. The structure of the Latent Variable Block for the teacher model is the same, except for the distribution of Posterior and Prior is Gaussian. 

\textbf{Posteriors.} The posterior distribution of $Z_i$ given $x$ and $z<i$ is defined as:
\begin{equation}
\label{eq:post}
\begin{aligned}
q_i &\triangleq U(\mu_i - \frac{1}{2}, \mu_i + \frac{1}{2})
\end{aligned}
\end{equation}
where $\mu_i$ is the output of the posterior branch, depending on the image $x$ and preceding latent variables $ z_{<i}$. Once $q_i$ is obtained, $z_i$ is sampled as $z_i \leftarrow \mu_i + u$, where $u$ is a random sample from $U(-\frac{1}{2}, \frac{1}{2})$ during training, and during testing, it is replaced with scalar quantization.

\textbf{Priors.} The prior distribution $p_i$ is defined as a conditional Gaussian convolved with a uniform distribution:
\begin{equation}
\label{eq:prior}
\begin{aligned}
p_i &\triangleq \mathcal{N}(\hat{\mu}_i, \hat{\sigma}_i^2) * U(- \frac{1}{2}, \frac{1}{2})
\end{aligned}
\end{equation}
where $\mathcal{N}(\hat{\mu}_i, \hat{\sigma}_i^2)$ represents the Gaussian probability density function. The mean $\hat{\mu}_i$ and standard deviation $\hat{\sigma}_i$ are predicted by the prior branch. The probability mass function (PMF) $P_i$ is then defined as:
\begin{equation}
P_i(n) \triangleq p_{i}(\hat{\mu}_i + n| z_{<i}), n \in \mathbb{Z}.
\end{equation}
which is used for the entropy coding/decoding of $z_i$.


\begin{figure}[htbp] 
    \centering 
    \includegraphics[width=0.75\linewidth]{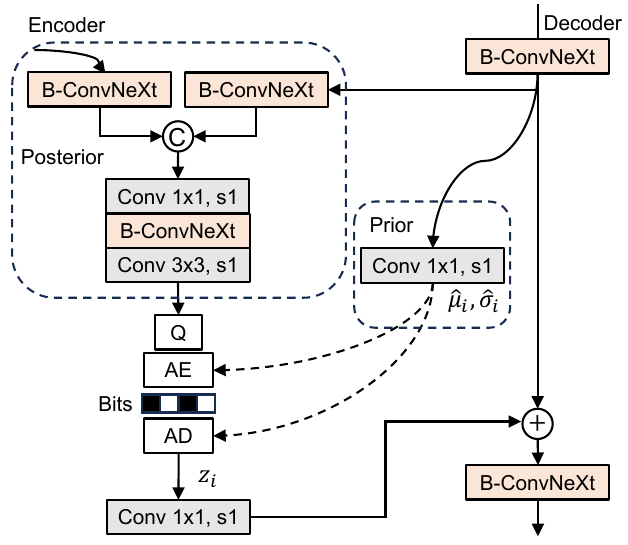}
    \vspace{-0.4cm}
    \caption{\textbf{Illustration of the $i$-th Latent Variable Block.}} 
    \vspace{-0.5cm}
    \label{fig:latent} 
\end{figure}

For details on the model architecture, please refer to the Supplementary Information.

\vspace{-0.3cm}
\section{Experiments}
\vspace{-0.3cm}

\subsection{Experimental Settings}
\vspace{-0.2cm}

\textbf{Training.} We use COCO2017~\cite{lin2014microsoft} as our training sets. It contains 118,287 images with a resolution of $640\times420$ pixels. We randomly cropped $256\times256$ patches for training. The model is trained for 2M iterations with a batch size of 32 and a learning rate of $2e^{-4}$. For all ablation experiments, we train the models for 500k
iterations. More details are shown in the supplementary information.  $B$ is pre-trained using the same settings and fixed during our BG-VAE training.


\textbf{Testing.} Three widely used benchmark datasets, including Kodak, Tecnick, and CLIC 2022, are used to evaluate the performance of the proposed method.

\vspace{-0.4cm}
\subsection{Quantitative Results}
\vspace{-0.2cm}

We compare our proposed method with prevalent NICs including fixed rate methods: M\&S Hyperprior~\cite{balle2018variational}, Cheng2020~\cite{cheng2020learned}, STF~\cite{zou2022devil},  ELIC~\cite{he2022elic}, TCM-S~\cite{liu2023learned}; variable rate method: QARV~\cite{duan2023qarv}; and rule-based method: VVC~\cite{bross2021overview}. We use VTM-18.0 All Intra as the anchor to calculate BD-Rate.

\begin{table*}[h]
    \centering
    \small
    \caption{Computational Complexity and BD-Rate Compared to Existing Learning-based Methods}
    \begin{tabular}{c|c|c|c|c|c|c|c|c}\hline\hline
        \multirow{2}{*}{Method} & \multirow{2}{*}{Total Params.}&\multicolumn{2}{c|}{Latency (CPU)} &\multicolumn{2}{c|}{Latency (GPU)}& \multicolumn{3}{c}{BD-Rate (\%) w.r.t. VTM 18.0}\\
        \cline{3-9}   & &Enc. &Dec.&Enc. &Dec.   & Kodak & Tecnick & CLIC2022  \\\hline
        M \& S Hyperprior~\cite{minnen2018joint} &98.4M &\textbf{0.759s} &0.830s & \textbf{0.033s} & \textbf{0.030s}&17.06\%&26.09\% &29.11\%\\
        Cheng2020~\cite{cheng2020learned} &115.3M& 3.605s&6.081s &1.048s  &2.376s &3.89\%&5.92\%&8.32\%\\
        STF~\cite{zou2022devil}   &599.1M&2.373s& 2.738s &  0.076s &0.068s&-2.55\%&-2.35\%&-1.17\% \\
        ELIC$^*$~\cite{he2022elic}  &202.8M& 2.131s&2.187s &  0.125s  & 0.062s&-5.69\%&-&- \\
        TCM-S~\cite{liu2023learned} &271.1M &\underline{0.837s} & 0.945s &0.095s&0.086s&-5.54\% &\underline{-8.15\%}&\textbf{-8.52\%} \\
        QARV~\cite{duan2023qarv}&\textbf{93.4M}&0.852s & \textbf{0.309s}  &0.098s&0.068s&\underline{-5.82\%}&-7.79\%&-6.13\%\\\hline
        BG-VAE  &\underline{97.4M} &0.990s & \underline{0.437s} &\underline{0.082s}&\underline{0.055s}&\textbf{-7.04\%}&\textbf{-8.21\%}&\underline{-6.33\%}\\\hline\hline
    \end{tabular}

\begin{tablenotes}
  \item {\bf Test Conditions}: Intel(R) Core(TM) i7-12700K CPU, Nvidia 3090 GPU. The enc./dec. time is averaged over all 24 images in Kodak, including entropy enc./dec. time. $^*$: We reproduced ELIC~\cite{he2022elic} to calculate the runtime. ELIC results on Kodak are officially provided.
  \item \textbf{Bold} and \underline{underlined} indicate the best and the second best, respectively.
  \vspace{-0.3cm}
  \end{tablenotes}
  \label{tab:bdrate}%
\end{table*}%
Table~\ref{tab:bdrate} reports the BD-Rate reduction of each method against the VVC anchor on three datasets. Our BG-VAE achieves impressive performance on each dataset, -7.04\% BD-Rate on Kodak, -8.21\% BD-Rate on Tecnick, and -6.33\% BD-Rate on CLIC 2022. Fig.~\ref{fig:mse_result} further plots RD curves of all methods. Our BG-VAE performs best on Kodak and Tecnick datasets and gains lower than TCM-S on CLIC2022. However, the total parameters of our BG-VAE are only 36\% of TCM-S, highlighting our high efficiency especially when considering our BG-VAE covers a wide range of bpp (0.2 bpp to 2.2 bpp).

\begin{figure*}[t] 
\newcommand{\mywidth}{0.3}
\centering 
\subfigure[Kodak]{\includegraphics[width=\mywidth\linewidth]{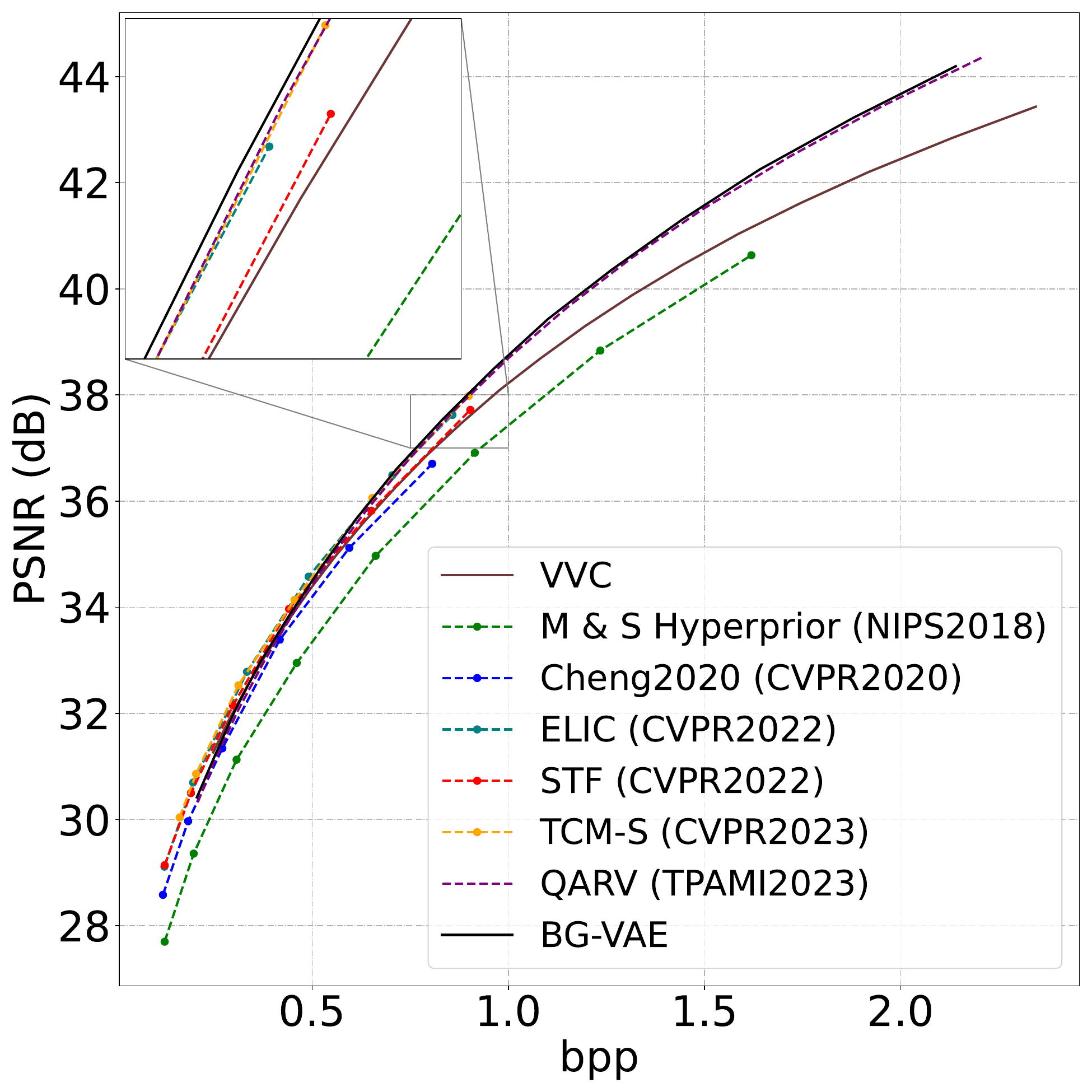}\label{subfig:kodak}} 
\subfigure[Tecnick 1200x1200]{\includegraphics[width=\mywidth\linewidth]{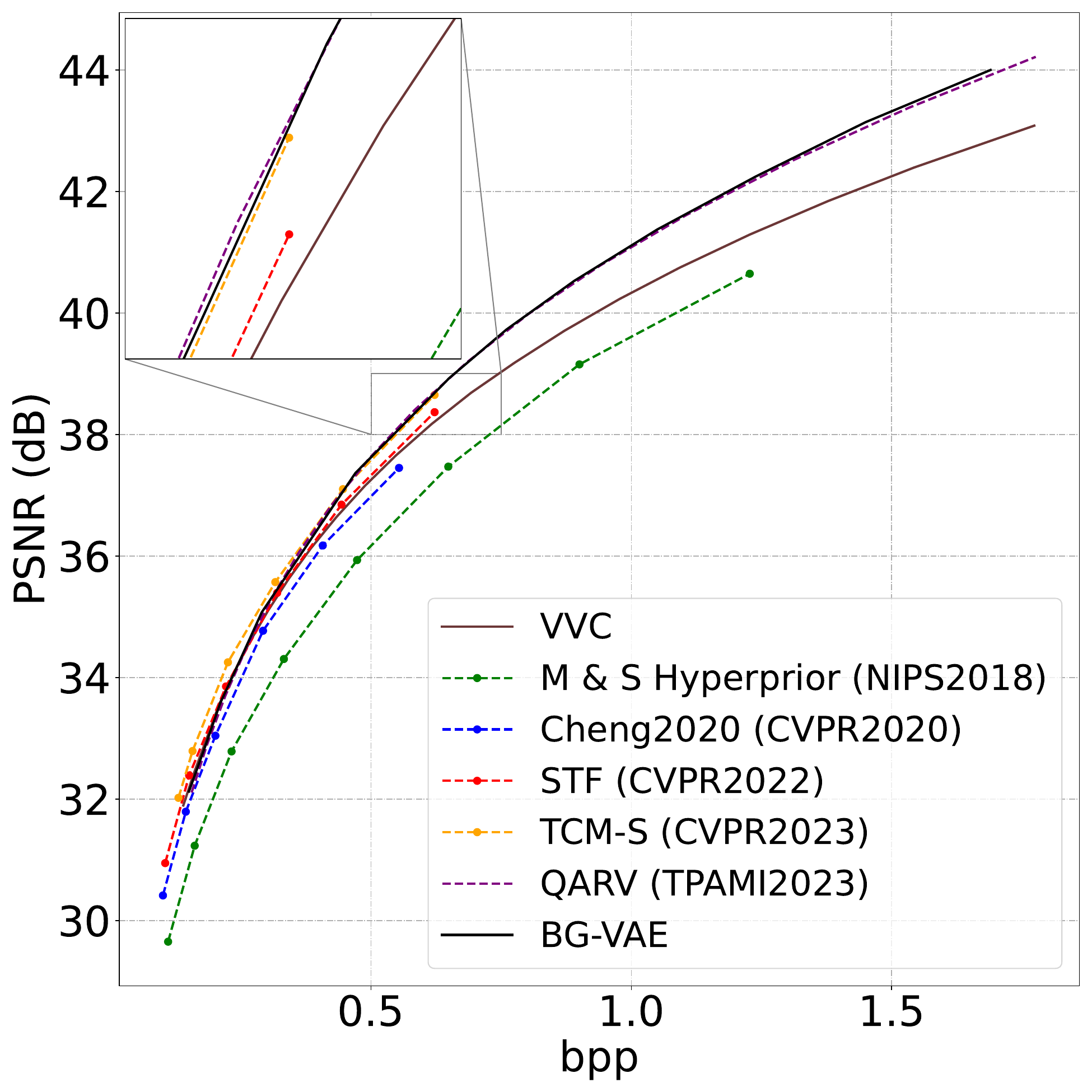}\label{subfig:Tecnick}}
\subfigure[CLIC 2022]{\includegraphics[width=\mywidth\linewidth]{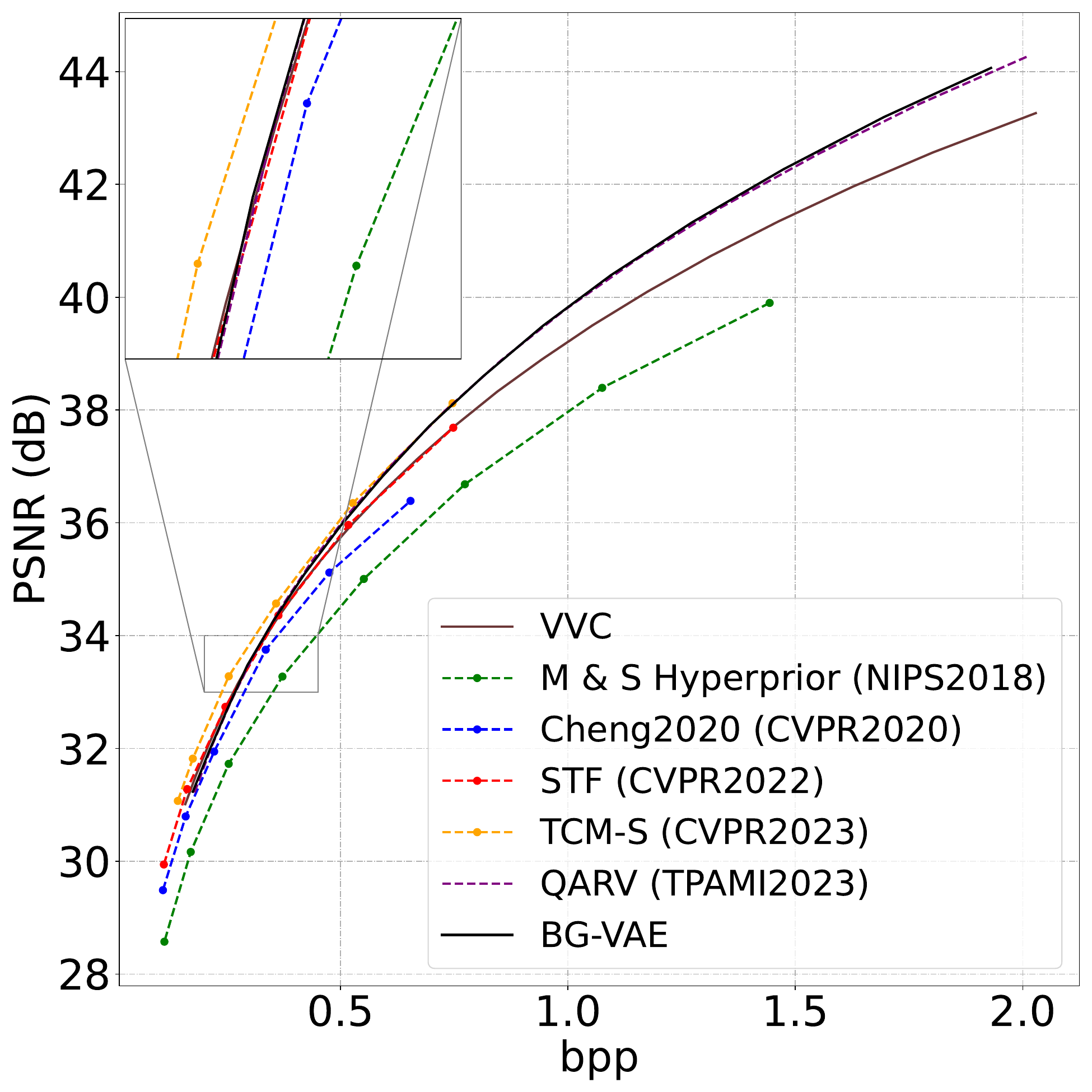}\label{subfig:CLIC}}
\vspace{-0.4cm}
\caption{\textbf{RD curves of various methods. }{\it Please zoom in for more details}.} 
\vspace{-0.5cm}
\label{fig:mse_result} 
\end{figure*}

\vspace{-0.3cm}
\subsection{Complexity}
\vspace{-0.2cm}

We measure the computational complexity using the total parameters (The total parameters of the fixed rate method are obtained by summing up model parameters for all bpp.), encoding time (Enc.), and decoding time (Dec.) on CPU and GPU, as shown in Table~\ref{tab:bdrate}.
In comparison to other methods, our model's total parameter is notably efficient, being the second smallest and exceeding QARV by only 4M parameters. Despite this minimal difference, our model achieves a 1\% higher BD-Rate Reduction. While M\&S Hyperprior's parameter is similar, its performance on the Kodak lags behind our results by 24\% BD-Rate. In contrast, the other methods have a substantially higher number of parameters. Regarding latency, BG-VAE's encoding time is comparable with other methods and the decoding time is only marginally longer than QARV on the CPU and longer than M\&S Hyperprior on the GPU, which demonstrates the high efficiency of BG-VAE.

\vspace{-0.3cm}
\subsection{Ablation Study}
\vspace{-0.2cm}

A series of ablation studies are conducted to verify the contribution of the bound-guided method and each module.

\textbf{Bound-Guided Training:}
One of the key components of our proposed method is the bound-guided training, as shown in Table~\ref{tab:Distillation}. Eliminating the bound-guided training, see ``BG-VAE base model", results in an average increase of 0.99\% BD-Rate compared to ``BG-VAE", highlighting the bound-guided training's effectiveness. Further, replacing the bound model with a larger teacher model (LM), as shown in ``W / LM", demonstrates the efficiency of using the bound to guide the NIC. Notably, due to the larger number of parameters of LM (124.5M vs 97.4M), training the LM incurs greater resource expenditure than the bound. Additionally, maintaining the bound model while replacing the $\mathcal{L}_\text{feature}, \mathcal{L}_\text{rs}$ with $\mathcal{L}_\text{MSE}$ (``W / $\mathcal{L}_\text{MSE}$") further underscores the affinity matrix's superiority.  In BG-VAE, features $F_1$ to $F_8$, depicted in Fig~\ref{subfig:model} and Section~\ref{sec:over}, are used to calculate $\mathcal{L}_\text{feature}$. It's also noteworthy that limited supervision, using only $F_5$ to $F_8$ (``W / $F_5$ to $F_8$"), proves less effective."

\begin{table}[h]
  \centering
  \caption{Ablation experiments on bound-guided framework}
  \resizebox{\linewidth}{!}   {
    \begin{tabular}{c|c|c|c}\hline\hline
    Settings      & {Kodak} & Tecnick & CLIC2022 \\\hline
    BG-VAE base model  & {-3.58\%} & {-3.88\%} &  -2.30\% \\
    W / LM  & {-3.79\%} & {-4.12\%} & -3.09\% \\
    W / $\mathcal{L}_\text{MSE}$ & {-3.97\%} & {-4.28\%} & -3.14\% \\
    W / $F_5$ to $F_8$  & {-3.92\%} & {-4.52\%} & -3.12\% \\\hline
    BG-VAE & {-4.46\%} & {-4.78\%} & -3.50\% \\\hline\hline 
    \end{tabular}}%
    \vspace{-0.3cm}
  \label{tab:Distillation}%
\end{table}%

\textbf{Network Architecture:}
we conducted individual assessments of three advanced modules in Table~\ref{tab:arh}. Removing Wavelet sampling (``W / o Wavelet sampling") had negligible impact on Kodak images but significantly affected high-resolution Tecnick and CLIC2022 images, likely due to their richer textural details, suggesting its importance in handling textural details due to the invertibility of Wavelet downsampling, unlike convolution downsampling which loss details. Additionally, eliminating the Balancing factor (``W / o Balancing factor") which entails using the ConvNeXt block in place of the B-ConvNeXt block, uniformly decreased performance across all datasets, underscoring its effectiveness with a simple design. Lastly, our experiment ``W / o Cross-Attn" demonstrates its integral role across all three datasets. This confirms the effectiveness of Cross-Attention in leveraging global information to enhance overall performance.

\begin{table}[h]
  \centering
  \caption{Ablation experiments on model architecture}
  \resizebox{\linewidth}{!}   {
    \begin{tabular}{c|c|c|c}\hline\hline
    Settings     & {Kodak} & Tecnick & CLIC2022 \\\hline
    W / o Wavelet sampling & {-3.53\%} & {-1.80\%} & -1.21\% \\
    W / o Balancing factor  & {-3.11\%} & {-2.45\%} & -1.15\% \\
    W / o Cross-Attn   & {-3.03\%} & {-3.09\%} & -1.79\% \\\hline
    BG-VAE base model & {-3.58\%} & {-3.88\%} & -2.30\% \\\hline\hline
    \end{tabular}}%
    \vspace{-0.3cm}
  \label{tab:arh}%
\end{table}%
\vspace{-0.5cm}


\section{Conclusion}
\vspace{-0.2cm}

In this paper, we present a theoretical Bound-Guided hierarchical VAE (BG-VAE) for NIC. The core of BG-VAE is its bound-guided training, which effectively boosts performance by utilizing the theoretical bound to guide the NIC model in training. Additionally, we develop several efficient modules that adeptly harness spatial-spectral information, forming the backbone of BG-VAE. These designs together present a versatile, variable-rate compression method. Extensive experimental results underscore BG-VAE's superior performance when compared to the traditional rule-based VVC and other neural image codecs, marking a significant advancement in the field of image compression.


\bibliographystyle{IEEEbib}
{
\vspace{-0.3cm}
\footnotesize
\bibliography{QARV2}
}

\clearpage

\fi
\end{document}